\documentclass{sig-alternate-NoCopyright-PageNb}
\usepackage[noadjust]{cite}
\usepackage{graphicx}
\usepackage{subfigure}
\usepackage{url}

\newcommand{\beq}{\begin{equation}}
\newcommand{\eeq}{\end{equation}}
\def\bearn{\begin{eqnarray*}}
\def\eearn{\end{eqnarray*}}
\def\barr{\begin{array}}
\def\earr{\end{array}} 

\def\bt{BitTorrent}
\def\p2p{P2P}
\def\P2p{P2P}

\newcommand{\urlsamefont}[1]
{
\urlstyle{same}\url{#1}
}

\begin{document}

\title{Pushing BitTorrent Locality to the Limit}

\numberofauthors{1}
\author{
\alignauthor Stevens Le Blond, Arnaud Legout and Walid Dabbous\\
\affaddr{I.N.R.I.A., France}
\vspace{3mm}
\email{Contact: \{stevens.le\_blond, arnaud.legout, walid.dabbous\}@inria.fr}
}

\maketitle 

\sloppy

\begin{abstract}
  Peer-to-peer (P2P) locality has recently raised a lot of
    interest in the community. Indeed, whereas P2P content
    distribution enables financial savings for the content providers,
    it dramatically increases the traffic on inter-ISP links.

    To solve this issue, the idea to keep a fraction of the P2P
    traffic local to each ISP was introduced a few years ago. Since
    then, P2P solutions exploiting locality have been
    introduced. However, several fundamental issues on locality still
    need to be explored. In particular, how far can we push locality,
    and what is, at the scale of the Internet, the
    reduction of traffic that can be achieved with locality?

    In this paper, we perform extensive experiments on a controlled
    environment with up to $10\,000$ BitTorrent clients to evaluate
    the impact of high locality on inter-ISP links traffic and peers
    download completion time.

    We introduce two simple mechanisms that make high locality
    possible in challenging scenarios and we show that we save up
    to several orders of magnitude inter-ISP traffic 
    compared to traditional locality without adversely impacting peers
    download completion time. In addition, we crawled $214\,443$
    torrents representing $6\,113\,224$ unique peers spread among
    $9\,605$ ASes. We show that whereas the torrents we crawled
    generated $11.6$ petabytes of inter-ISP traffic, our locality policy
    implemented for all torrents would have reduced the global
    inter-ISP traffic by $40\%$.

\end{abstract}

\section{Introduction}
\label{intro}

Content distribution is today at the core of the services provided by
the Internet. However, distributing content to a large audience is
costly with a classical client-server or CDN solution. This is the
reason why content providers start to move to \p2p{} content
distribution that enables to significantly reduce their cost without
penalizing the experience of users. One striking example is iPlayer, a
\p2p{} service for video-on-demand that distributes recent BBC
programs.

However, whereas current \p2p{} content distribution solutions like
BitTorrent are very efficient, they generate a huge amount of traffic
on inter-ISP links. Indeed, in BitTorrent, each peer that downloads a given
content is connected to a small subset of peers picked at random among
all the peers that download that content. In fact, even though peers in
the same ISP are downloading the same content they are not necessarily
connected to each other. As a consequence, peers unnecessarily
download most of the content from peers located outside of their ISP.

Therefore, even if current \p2p{} content replication solutions
significantly reduce content provider costs, they cannot be promoted
as a global solution for content replication as they induce huge costs
for ISPs. In particular, the current trend for ISPs is to block \p2p{}
traffic~\cite{blocking}.

One solution to this problem is to use \p2p{} locality, that is to constrain
\p2p{} traffic within ISPs' boundaries in order to minimize the amount of
inter-ISP traffic.

The seminal work of Karagiannis et al. \cite{kara05_IMC} is the first
one to suggest the use of locality in a \p2p{} system in order
to reduce the load on inter-ISP links. They show on real traces the
potential for locality (in particular spatial and temporal correlation
in the requests for contents) and, based on simulation on a BitTorrent
tracker log, they evaluate the benefit of several architectures and in
particular a \p2p{} architecture exploiting locality. 
More recently, Xie et al. \cite{p4p} proposed P4P, an architecture to
enable cooperation between \p2p{} applications and ISPs. They show
by performing large field tests that P4P enables reduction of external
traffic for a monitored ISP and enables a reduction on the peers
download completion time. 
Choffnes et al. \cite{ono} proposed Ono, a BitTorrent extension that
leverages on a CDN infrastructure to localize peers in order to group
peers that are close to each other. They show the benefit of Ono in
terms of peers download completion time and suggest, using indirect
measurements (IP hops and AS hops among peers), that Ono can also
reduce inter-ISP traffic.

With those works, there is no doubt that \p2p{} locality has some
benefits and that there are several ways to implement it. However, two
fundamental questions are left unanswered by those previous works.
\begin{itemize}
\item \textit{How far can we push locality?} In all proposed solutions the
  number of inter-ISP connections is kept high enough to guarantee a
  good robustness to partitions, i.e., a lack of connectivity among
  set of peers resulting in a poor download completion time. However,
  this robustness is at the expense of a larger inter-ISP traffic. How
  far can we push locality without impacting the robustness to
  partition of the \p2p{} protocol?

\item \textit{What is, at the scale of the Internet, the reduction of
    traffic that can be achieved with locality?} It might be argued
  that \p2p{} locality will bring little benefits at the scale of the
  Internet, in case most ISPs have just a few peers, thus few
  potential benefits with peers locality. Therefore, the question is,
  what is the distribution of peers per ISP in the Internet, and what
  would be the inter-ISP bandwidth savings achieved with a locality
  policy.  Previous works looking at inter-ISP bandwidth savings
  either consider indirect measurements (like the distribution of the
  number of AS between connected peers), partial measurements (like
  the monitoring of a specific ISP), or simulations (like comparing
  various content distribution scenarios based on the location of
  peers obtained from a tracker log). For instance, Xie et
  al. \cite{p4p} reported results on inter-ISP
  savings with P4P for a \textit{single} ISP.
\end{itemize}

  The answers to those questions are fundamental if ever \p2p{}
  content replication is used by content providers for large scale
  distribution. In that case, it is likely that ISPs will need to know
  the amount of inter-ISP traffic they can save with locality, and
  that they will request content providers to minimize this traffic due
  to \p2p{} applications accordingly. At the same time, the
  content providers will need a clear understanding of the impact of
  this reduction of traffic on their customers.

Our contribution in this paper is to answer those questions by running
extensive large scale \bt{} experiments (with up to $10\,000$ real
BitTorrent clients) in a controlled environment, and by using real
data we crawled in the Internet on $214\,443$ torrents representing
$6\,113\,224$ unique peers spread among $9\,605$ ASes. Our
work can be summarized with the two following key contributions.

i) We show that we can push \bt{} locality much further than what was
previously proposed, which enables to reduce by several orders of
magnitude the inter-ISP traffic and to keep the peers download
completion time low. In particular, we show on experiments including
real world data that the reduction of inter-ISP traffic and the peers
download completion time are not significantly impacted by the torrent
size, the number of peers per ISP, and the churn. Finally, we propose
new strategies to improve the efficiency and robustness of our
locality policy on challenging scenarios defined from real world
torrents. 

ii) We show that at the scale of the $214\,443$ torrents we crawled,
ISPs can largely benefit from locality. In particular, whereas all the
torrents crawled generated $11.6$ petabytes of inter-ISP traffic, high
locality would have saved $40\%$, i.e., $4.6$ petabytes, of inter-ISP
traffic. This results is significantly different from the inter-ISP
bandwidth savings reported by Xie et al. \cite{p4p}. Indeed, they
reported a reduction of inter-ISP traffic with P4P around $60\%$, but
for a single ISP with a single large torrent. Thus, they did not
evaluated the reduction of \bt{} traffic at the scale of the Internet,
but for a single ISP. The result we report is an estimation for
$214\,443$ real torrents spread across $9\,605$ ASes, thus capturing
 the variety of torrent sizes and distribution of peers per AS we
can find in the Internet.

The remaining of this paper is organized as follows. We define
the locality policy we use for our evaluation in
section~\ref{sec:method-locality}, then we describe our experimental
setup, and define metrics in section~\ref{method}. We discuss the
impact of the number of inter-ISP connections in
section~\ref{locality} and focus on a small number of inter-ISP
connections in section~\ref{small}.  In
section~\ref{sec:real-world-scenarios}, we present results obtained
from a large crawl of torrents in the Internet. In section~\ref{work},
we discuss the related work. Finally, we conclude in
section~\ref{sec:conclusion}.

\section{Locality Policy}
\label{sec:method-locality}
In this paper, we make a experimental evaluation of the two questions
discussed in the introduction. To do so, we introduce in the following
a locality policy that we use to perform our evaluation. We do not
claim our locality policy to be a definitive solution that should be
deployed. Instead, it is a simple implementation that we used for our
evaluation. Yet, we identified two important strategies that we
recommend to consider, even in a modified form, for any implementation
of a locality policy.

In the following, we refer to \textit{\bt{} policy} when the
tracker does not implement our locality policy, but the regular random
policy.

\subsection{Implementation of the Locality Policy}
\label{sec:defin-local-policy}

We say that a connection is \textit{inter-ISP} when two peers in two different
ISPs have established a direct \bt{} connection, and that it is \textit{intra-ISP}
when the two peers are from the same ISP.  The goal of our
\textit{locality policy} is to limit the number of inter-ISP
connections, the higher the locality, the smaller the
number of inter-ISP connections.

We say that an inter-ISP connection is \textit{outgoing}
(resp. \textit{incoming}) for an ISP if the connection was initiated
by a peer inside (resp. outside) this ISP. However, once a connection
is established it is fully bidirectional.

In order to control the number of inter-ISP connections, we assume
that the tracker can map each peer to its ISP. How this mapping is
performed is orthogonal to our work. For instance, the tracker can
simply map peers to ASes using precomputed mapping information
obtained from BGP tables \cite{routeviews}. In case the AS level is
not appropriate for ISPs, the tracker can use more sophisticated
information as the one offered by, for instance, the P4P
infrastructure \cite{p4p}.

The only one parameter of our locality policy is the \textit{maximum
  number of outgoing inter-ISP connections per ISP}. The tracker
maintains for each ISP the number of peers outside this ISP that it
returned to peers inside, along with the identity of the peers
inside. This way the tracker maintains a reasonable approximation of
the number of outgoing inter-ISP connections for each ISP. When a peer
$P$ asks the tracker for a new list of peers, the tracker will: map
this peer to the ISP $I_p$ it belongs to; return to this peer a list
of peers inside $I_p$; if the maximum number of outgoing inter-ISP
connections per ISP is not yet reached for $I_p$, return one
additional peer $P_o$ outside $I_p$ and increment by one the counter
of the number of outgoing connections for $I_p$. We also add a
randomization factor to distribute the outgoing connections evenly among
the peers of each ISP.

Each regular \bt{} client contacts periodically, typically every $30$ minutes,
the tracker to return statistics. Each time a peer leaves the torrent,
it contacts the tracker so that it can remove this peer from the list
of peers in the torrent. In case the client does not contact the tracker
when it leaves (for instance, due to a crash of the client), the
tracker will automatically remove the peer after a predefined period
of time, typically $45$ minutes, after the last connection of the peer
to the tracker. Our locality policy uses this information to maintain
an up-to-date list of the number of outgoing inter-ISP connections per
ISP. 

When the tracker implements our locality policy, it applies the
locality policy to all peers except the initial seed. Because the goal
of the initial seed is to improve diversity, the tracker selects the
neighbors of the initial seed using the \bt{} policy. However, we
apply the locality policy to all the other seeds, that is all the
leechers that become seed during the experiments. Note that the
traffic generated by the initial seed is negligible compared to the
aggregated traffic of the torrent.

\subsection{Round Robin Strategy}
\label{sec:round-robin-strategy}
Our locality policy controls the number of outgoing inter-ISP
connections per ISP. When a peer $P$ from the ISP $I_p$ opens a new
connection to a peer $P_o$ from the ISP $I_{p_o}$, the connection is
outgoing for $I_p$, but incoming for $I_{p_o}$. As both outgoing
and incoming connections account for the total number of inter-ISP
connections, it is important to define a strategy for the selection of
peer $P_o$ returned by the tracker to peer $P$.

We define two strategies to select this peer $P_o$.  The first
strategy, the default one, consists in selecting $P_o$ at random among
all peers outside $I_p$. While this strategy is straightforward, it
has the notable drawback that the largest ISPs have a higher
probability to have a peer selected than other ones. Therefore, large
ISPs will have more incoming connections than small ones. Thus, it is
likely that in this case, as connections are bidirectional, the
inter-ISP traffic will be higher for large ISPs (we confirm this
intuition in section~\ref{sec:impact-locality-real}). In the second
strategy that we call Round Robin (RR), the tracker selects first the
ISP with a round robin policy and then selects a peer at random in the
selected ISP. This way, the probability to select a peer in a given ISP
is independent of the size of this ISP.  

In scenarios with a same number of peers for each ISP, both strategies
are equivalent. Therefore, as all the experiments in
section~\ref{locality} and~\ref{small} consider an homogeneous
number of peers for each ISP, we only present the results with the
default strategy. We perform a detailed evaluation of the RR strategy
in section~\ref{sec:real-world-scenarios}.

\subsection{Partition Merging Strategy}
\label{sec:part-merg-strat}
One issue with a small number of inter-ISP connections is the higher
probability to have partitions in the torrent. Indeed, if peers who
have inter-ISP connections leave the torrent and no new peer joins the ISP, then
this ISP will form a partition. In order to repair partitions we
introduce an additional strategy called Partition Merging (PM)
strategy. The problem of partition in \bt{} is not specific to our
locality policy, but any locality policy favors its apparition.

The implementation of the Partition Merging strategy is the
following. On the client side, each leecher monitors the pieces
received by all its neighbors using the regular \bt{} HAVE
messages. If during a period of time randomly selected in $[0,T]$,
with $T$ initialized to $T_0$, the leecher cannot find any piece it
needs among all its neighbors (i.e., each neighbor has a subset of the
pieces of the leecher), it recontacts the tracker with a PM flag,
which means that the leecher believes there is a partition and that it
needs a connection to a new peer outside its ISP. In case the tracker
does not return a new peer, or if after receiving this new peer the
leecher does not observe any new piece it needs, the leecher performs
an exponential backoff of $T$, that is $T \leftarrow T*2$. As soon as
the leecher sees among its neighbors a piece it needs, it resets $T$
to $T_0$. This backoff reduces the load on the tracker but does
not prevent an implosion of requests at the tracker in case of very
large torrents. This issue, known as the \textit{feedback implosion
  problem} in the literature, can be solved using several techniques
\cite{312251}. However, a detailed description of a feedback implosion
mechanism for the PM strategy is beyond the scope of this paper.

On the tracker side, the tracker maintains for each ISP a flag that
indicates whether it answered a request from a peer with the PM flag
within the last $T_1$ minutes, i.e., the tracker returned to a peer of
this ISP a peer outside. The tracker will return at most one peer
outside each ISP every $T_1$ minutes in order to avoid exploiting this
strategy to bypass the locality policy.

The detailed evaluation of the impact of the initial values of the
timers is beyond the scope of this study. The choice of the values is
a tradeoff between reactivity and erroneous detection of
partitions. In this study, we set $T_0$ and $T_1$ to one minute, and
we show that it efficiently detects partitions without significantly
increasing the inter-ISP traffic.

This strategy might be abused by an attacker. Indeed, as the PM
strategy detects partitions relying on the accuracy of the HAVE
messages sent by neighbors, an attacker might generate dummy HAVE to
prevent peers of an ISP to detect a partition. However, this is not an issue in the context
of our study, as we work on a controlled environment, without
attackers. In addition, we don't
believe this is a major issue for the following two reasons. First, an
attacker must be a neighbor of all the peers of an ISP to attack
it. However, with the locality policy, the attacker must be in the ISP
it wants to attack, otherwise it has a very low probability to become
one of the ISP's peers neighbor. That makes the attack hard to deploy at the
scale of a torrent. Second, instead of relying on the monitoring of
HAVE messages, a peer can rely on pieces it receives. For instance, a
peer can combine the current PM strategy with the additional criterion
that it also generates a PM request to the tracker in case it does not
receive any new piece within a $5$ minutes interval. It is beyond the
scope of this study to perform a detailed analysis of variations of
the PM strategy, which has to be addressed in future work.

As this strategy has no impact on our experiments when there is no
partition, we present results in section~\ref{locality}
and~\ref{small} without the PM strategy unless explicitly specified,
that is when there is a partition and that the PM strategy changes the
result. We perform a detailed analysis of the PM strategy in
section~\ref{sec:real-world-scenarios}.

\subsection{Granularity of the Notion of Locality}
Our locality policy is designed to keep traffic local to
ISPs. However, we are not restricted to ISPs, and our locality policy
can keep traffic local to any network region as long as the tracker is
aware of the regions and has a mean to map peers to those regions. For
instance, a tracker can use information offered by a dedicated
infrastructure like the P4P infrastructure \cite{p4p}. In particular,
when we focus on real world scenarios in
section~\ref{sec:real-world-scenarios}, we will use ASes instead of
ISPs.

\section{Methodology}
\label{method}
In this section, we describe our experimental setup, and
the metrics that we consider to evaluate our experiments.

\subsection{Experimental Setup}
\label{sec:experimental-setup}
In this paper, we have run large scale experiments to evaluate the
impact of our locality policy on inter-ISP traffic and \bt{} download completion
time. We have run experiments instead of simulations for two main
reasons. First, it is hard to run realistic (packet level discrete)
\p2p{} simulations with more than a few thousand of peers due to the
large state generated by each peer and the packets in transit on the
links. Moreover, at that scale, simulations are often slower than
real time. Second, the dynamics of \bt{} is subtle and not yet deeply
understood. Running simulations with a simplified version of \bt{} may
hide fundamental properties of the system.

As we will see during the presentation of our results, we observe
behaviors that can only be pointed out using real experiments at large
scale, with up to $10\,000$ peers. 

We now describe the experimentation platform on which we run all our
experiments, the BitTorrent client that we use in our experiments, and
how we simulate an inter-ISP topology on top of the platform.

\subsubsection{Platform}
\label{sec:platform}

We obtain all our results by running large scale experiments with a
real BitTorrent client.

We run all our experiments on a dedicated experimentation platform.  A
typical node in this platform has bi or quad-core AMD Opteron CPU, $2$
to $4$GB of memory, and a gigabit Ethernet connectivity. The platform
we used consists of $178$ nodes. Once a set of nodes is reserved, no
other experiment can run on parallel on those nodes. In particular,
there is no virtualization on those nodes. Therefore, experiments are
totally controlled and reproducible.

The BitTorrent client used for our experiments is an instrumented
version of the mainline client \cite{btClientInstru}, which is based
on version $4.0.2$ of the official client \cite{bit_site}. This
instrumented client can log specific messages received and
sent. Unless specified otherwise, we use the default parameters of
this client. In particular, each peer uploads at $20$kB/s to $4$ other
peers, and the maximum peer set size is $80$. We will vary the upload
capacity when studying the impact of heterogeneous upload capacities
in section~\ref{locality} (see section~\ref{sec:exper-param-locality}
for a description of our heterogeneous scenario). We also use the
choke algorithm in seed state of the official client in its version
$4.0.2$. This algorithm is somewhat different, as it is fairer and
more robust than the one implemented in most BitTorrent clients
today. However, as it only impacts the seed, we do not believe this
algorithm to have a significant impact on our results.

Our client does not implement a gossiping strategy to discover peers,
like Peer Exchange (PEX) used in the Vuze client. Whereas it is
possible to make PEX locality aware, it is beyond the scope of this
study to make a detailed discussion of this issue.

We use the following default parameters for our experiments, unless
otherwise specified. Peers share a content of $100$MB that is split
into pieces of $256$kB. By default, all peers including the initial
seed start within the first $60$ seconds of the experiments. However,
we will also vary this parameter in
section~\ref{sec:evaluation-churn-real} when studying the impact of
churn (see section~\ref{sec:evaluation-churn-real} for a description
of our scenario with churn). Once a leecher has completed its download,
it stays $5$ minutes as seed and then leaves the torrent. We have
chosen $5$ minutes in order to give enough time for peers to upload
the last pieces they have download before becoming a seed.  However,
it should not impact our results because $5$ minutes is small compared
to the optimal download completion time ($83$ minutes). The initial
seed stays connected for the entire duration of the experiment.

We run all our experiments with up to $100$ BitTorrent clients per
physical node. Therefore, for torrents with $100$, $1\,000$, and $10\,000$
peers, we use respectively $1$, $10$, and $100$ nodes for the
leechers, plus one node for the seed and the tracker. Each client on a
same node uses a different port to allow communication among those
clients. We have performed a benchmarking test to find how many
clients we can run on a single node without a performance penalty that
we identify with a decrease in the client download time for a
reference content of $100$MB. We have found that we can run up to
$150$ clients uploading at $20$KB/s on a single node without
performance penalty. 
To be safe, we run no more than
$100$ clients uploading at $20$kB/s on one node, or $2$MB/s of
BitTorrent workload. 
When we
will vary the upload capacity of clients in section~\ref{locality}, we
will then adapt the number of clients per node so that the aggregated
upload capacity per node is never beyond $2$MB/s.

\subsubsection{Inter-ISPs Topology}
\label{sec:isps-topology}

We remind that our goal is to evaluate the impact of the number of
inter-ISP connections on inter-ISP traffic and peers
performance. Therefore, we simulated an inter-ISP topology on top of
the experimentation platform we use to run our experiments. We explain,
in the following, how we simulated this topology and how representative
it is of the real Internet.

For all our experiments, we assume that we have a set of stub-ISPs
that can communicate among each other. On top of this topology, we
consider two scenarios. The first scenario is when all stub-ISPs have
a single {\it peering} link to each other, thus the topology of the
network is a full mesh. We refer
to a peering link as a link for which an ISP does not pay for
traffic. However, the peering technology is expensive to upgrade so
ISPs are interested in reducing the load on those links. The second scenario is
when each stub-ISP is connected with a \textit{transit} link to a {\it single}
transit-ISP. All peers are in stub-ISPs. Therefore, there is no
traffic with a source or a destination in the transit-ISP. We refer to
a transit link as a link on which traffic is billed according to the
$95$-th percentile. Therefore, ISPs are interested in reducing the
bursts of traffic on those links.

We observe that both scenarios are simply a different interpretation of
a same experiment, as all peers are in stub-ISPs and the traffic flows
from one stub-ISP to another one. In the following, we refer to {\it
  inter-ISP link} when our discussion applies to both peering and
transit links.

In our experiments, the notion of ISPs and inter-ISP links is virtual,
as we run all our experiments on an experimentation platform. To simulate the
presence of a peer in a given ISP, before each experiment, we create a
static mapping between peers and ISPs. We use this mapping to compute
offline the traffic that is uploaded on each inter-ISP link of the
stub-ISPs. For instance, imagine that peer $P_A$ is mapped to the ISP
$A$ and peer $P_B$ is mapped to the ISP $B$. All the traffic sent from
$P_A$ to $P_B$ is considered as traffic uploaded by the ISP $A$ to the
ISP $B$ with a peering link in the first scenario or with a transit
link via the transit ISP in the second scenario.

Our experiments are equivalent to what we would have obtained in the
Internet with real ISPs and inter-ISP links except for latency.  We
argue that latency would not significantly change our results because:
i) we limit the upload capacity on each \bt{} client, thus the RTT is
not the limiting factor for the end-to-end throughput; ii) the choking
algorithm is insensitive to latency by design, as \bt{} computes the
throughput of neighbors (used to unchoke them) over a $10$ seconds
interval, which should alleviate the impact on \bt{} of the TCP ramp
up~\cite{COHE03_WEP2P} due to latency.

We experiment with and without bottlenecks in the network. By default,
there is no bottleneck in the network because the aggregated traffic
generated by our experiments is always significantly lower than the
bottleneck capacity of the experimentation platform. However, we also create artificial
bottlenecks on inter-ISP links to evaluate their impact on inter-ISP
traffic and performances (see section~\ref{bottleneck} for the
description of how we limit the inter-ISP link capacity). It is important to
experiment the impact of bottlenecks on inter-ISP links because the
choking algorithm selects peers according to their throughput.
Therefore, bottlenecks may significantly change BitTorrent's behavior.

Finally, we have not considered a hierarchy of transit-ISPs. We show
in section~\ref{sec:estim-local-benef-real} that there is a huge
amount of inter-ISP traffic generated by \bt{}. Even if the proposed
locality policy already significantly reduces this traffic,
optimizations for the transit-ISPs still makes sense. We keep the
detailed evaluation of the optimization of the traffic in a hierarchy
of transit ISPs for future work.

\subsection{Evaluation Metrics}
\label{sec:evaluation-metrics}

To evaluate our experiments, we consider three metrics: the content
replication overhead, the $95$th percentile, and the peer slowdown.

\begin{description}
\item[\textbf{Overhead}] 
For each stub-ISP, we monitor the number of pieces
that are uploaded from this stub-ISP to any other stub-ISP during the
experiment. Then, to obtain the per-ISP content replication overhead,
we normalize the amount of data uploaded by the size of the content
for the experiment. Thus, we obtain the overhead in unit of contents
that crosses an inter-ISP link. We call this metric the content
replication overhead, or overhead for short, because with the
client-server paradigm, ISPs with clients only would not upload any
byte. We use the overhead as a measure of load on peering links.

\item[\textbf{95th Percentile}] 
To obtain the $95$th percentile of the
  overhead, we compute the overhead by periods of $5$ minutes and then
  consider the $5$ minutes overhead corresponding to the $95$th
  percentile. The $95$th percentile is the most popular charging model
  used on the Internet \cite{Odlyzko00internetpricing}.

\item[\textbf{Slowdown}] 
 We define the ideal completion time of a peer
  as the time for this peer to download the content at a speed
  equivalent to the average of the maximum upload capacity of all peers. This is the
  best completion time, averaged over all peers, that can be achieved
  in a \p2p{} system in which each peer always uploads at its maximum
  upload capacity. The slowdown is the experimental peer download
  completion time normalized by the ideal completion time.  For
  instance, imagine that all peers have the same maximum upload
  capacity of $20$kB/s. An average peer slowdown of $1$ for $10\,000$
  peers means that there is an optimal utilization of the peers upload
  capacity, or that the peers are, on average, as fast as a
  client-server scenario in which we have $10\,000$ servers, one
  server per client sending at $20$kB/s.
\end{description}

\section{Impact of the Number of Inter-ISP Connections}
\label{locality}

The goal of this section is to explore the relation between the number
of inter-ISP connections and the overhead and slowdown. In
particular, we will evaluate how far we can push locality (that is,
how much we can reduce the number of inter-ISP connections)
to obtain the smallest overhead attainable and what is the impact of
this reduction on the slowdown.

\subsection{Experimental  Parameters}
\label{sec:exper-param-locality}
For this series of experiments, we set the torrent size to $1\,000$
peers, the number of ISPs to $10$, and the content size to $100$
MB. Therefore, there are $100$ peers per ISP in all the experiments of
the first series. To analyze the impact of the number of inter-ISP
connections on \bt{}, we then vary the number of outgoing inter-ISP
connections between $4$ and $40$ by step of $4$, and between $400$ and
$3600$ by steps of $400$. As we consider, in this section, scenarios
with the same number of peers for each ISP, the total number of
inter-ISP connections per ISP will be on average twice the number of
outgoing inter-ISP connections. We run experiments for each of
the three following scenarios.

\begin{description}

\item[\textbf{Homogeneous scenario with a slow seed}] 
 In this scenario
  both the initial seed and the leechers can upload at a maximum rate of
  $20$kB/s. As we have mentioned earlier, we run $100$ leechers per
  node, and we run the initial seed and the tracker on an additional
  node. According to the definition of locality policy from
  section~\ref{sec:defin-local-policy}, each peer has the same
  probability to have a connection to the initial seed, whichever ISP
  it belongs to. For instance, as the initial seed has a peer set of
  $80$, with $10$ ISPs, each ISP has in average $8$ peers with a
  connection to this initial seed.

\item[\textbf{Heterogeneous scenario}] 
We experiment with leechers
  with heterogeneous upload capacities and a fast initial seed.  In
  each ISP, one third of the peers uploads at $20$kB/s, one third
  uploads at $50$kB/s, and one third uploads at $100$kB/s. For
  simplicity, we run all the leechers with the same upload capacity on
  the same node. Because we have determined that the hard drives
  cannot sustain a workload higher than $2$MB/s, we run only $20$
  clients per node. For BitTorrent to perform optimally, the initial
  seed uploads at $100$kB/s, as fast as the fastest leechers. Each
  peer has the same probability to have a connection to the initial
  seed, whichever ISP it belongs to.

  We experiment with heterogeneous upload capacities for three
  reasons. The first reason is that non-local peers may be faster than
  local ones so the local peers may unchoke inter-ISP connections
  more often than intra-ISP connections, thus making the reduction of
  the number of inter-ISP connections inefficient to reduce inter-ISP
  traffic. The second reason is that local peers may be faster than
  non-local ones so inter-ISP connections may be rarely used to
  download new pieces, thus degrading performances. The third reason
  is that in case of heterogeneous upload capacities inside an ISP, if
  fast peers are those with the inter-ISP connections, slower peers
  may not be given pieces to trade among themselves, also degrading
  performances.

\item[\textbf{Homogeneous scenario with a fast seed}] 
 We experiment
  with leechers that upload at $20$kB/s and an initial seed that uploads at
  $100$kB/s. We run this additional experiment in order to understand
  whether the results obtained with the heterogeneous scenario are due
  to the fast initial seed or due to the heterogeneous capacities of leechers.

\end{description}

First, we evaluate the impact of the number of inter-ISP connections on overhead and $95$th
percentile. Then, we evaluate the impact of the number of inter-ISP connections on slowdown.

\subsection{Impact on Overhead}
\label{sec:locality-impact-overhead}

\begin{figure}[!t]
\centering
\includegraphics[width=1.0\columnwidth]{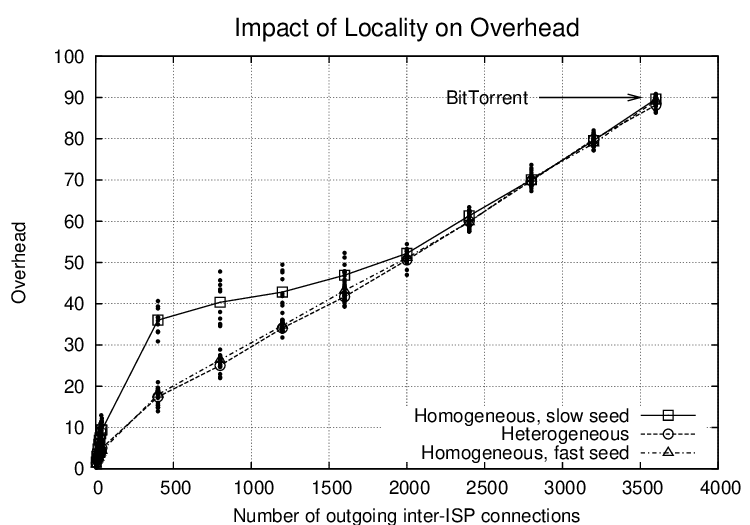}
\caption{\small{Overhead with $1\,000$ peers and $10$ ISPs. Each
    square, circle and triangle represents the average overhead on all
    ISPs in a given scenario. Each dot represents this overhead for
    one ISP.
}}
\label{overhead-locality}
\end{figure}

We observe in Fig.~\ref{overhead-locality} that for the two scenarios
with a well provisioned initial seed, i.e., the homogeneous fast seed
and the heterogeneous scenarios, the overhead increases linearly with
the number of outgoing inter-ISP connections. 
Indeed, when there is no congestion in the network and a uniform
repartition of the upload capacity of peers in each ISP, the
probability to unchoke a peer outside his own ISP is linearly
dependent on the number neighbors this peer has outside his own ISP,
thus it is linearly dependent on the number of outgoing inter-ISP
connections. We evaluate the impact of network bottlenecks in
section~\ref{bottleneck}.

The BitTorrent arrows in Fig.~\ref{overhead-locality}
and~\ref{slowdown-locality} represent the value 
of respectively overhead and slowdown achieved by BitTorrent
in the same scenario. Indeed, with $1\,000$ peers and $10$ ISPs of
$100$ peers, each peer has $10\%$ of connections inside his own ISP
with the \bt{} policy. Therefore, with \bt{} each ISP will have
$7\,200$ inter-ISP connections, $3\,600$ of those connections being
outgoing. Thus \bt{} corresponds to the case with $3\,600$ outgoing
inter-ISP connections in our experiments.

For all three scenarios, our locality policy enables to reduce by up to two
orders of magnitude the traffic on inter-ISP links. Indeed, we see in
Fig.~\ref{overhead-locality} that for $3\,600$ outgoing inter-ISP connections,
the case of the \bt{} policy, the overhead is close to $90$, and for
$4$ outgoing inter-ISP connections the overhead is close to $1$ for all three
scenarios. 

Surprisingly, we observe in Fig.~\ref{overhead-locality} that between
$400$ and $2\,000$ outgoing inter-ISP connections, there is a higher
overhead for the homogeneous scenario with a slow seed than for the
two other scenarios with a fast seed. Indeed, as there is a lower
piece diversity with a slow seed, peers in a given ISP will have to use more their inter-ISP
connections, thus a higher overhead, in order to download pieces that
are missing in their own ISP. We do not observe the same issue with
 a fast seed because this fast initial seed is
fast enough to guarantee a high piece diversity even for a small
number of outgoing inter-ISP connections.

\begin{figure}[!t]
\centering
\includegraphics[width=1.0\columnwidth]{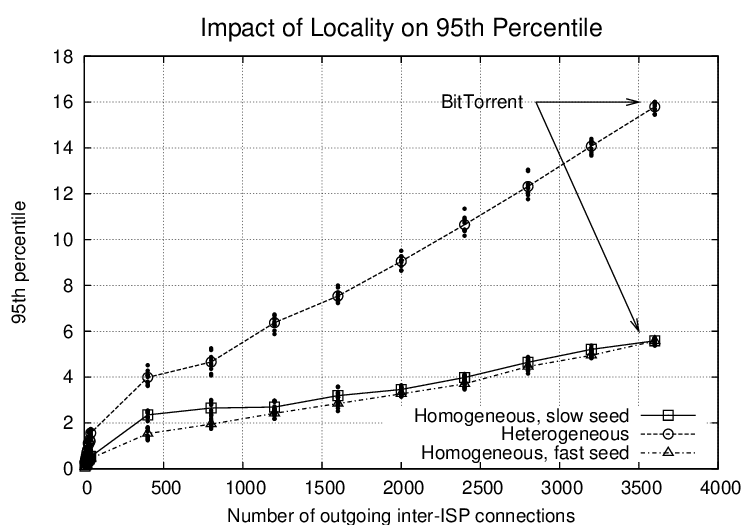}
\caption{\small{$95$th percentile with $1\,000$ peers and $10$
    ISPs. Each square, circle and triangle represents the average
    $95$th percentile on all ISPs for a given scenario. Each dot
    represents this $95$th percentile for one ISP.
}}
\label{peak-locality}
\end{figure}

We also observe a linear relation between the number of outgoing inter-ISP
connections and the $95$th percentile as well as a significant
reduction of the $95$th percentile for a small number of outgoing inter-ISP
connections in Fig.~\ref{peak-locality}. However, we observe that the
$95$th percentile for the heterogeneous scenario is much larger than
for the two other scenarios. This is because in the heterogeneous
scenario there are two third of the peers that are faster than $20$
kB/s, which is the upload capacity of all the peers for the two other
scenarios. Therefore, we see that even if the total amount of traffic
crossing inter-ISP links is not significantly impacted by the
distribution of the upload capacity of peers (see
Fig.~\ref{overhead-locality}), this distribution might have a major
impact on the $95$th percentile that is used for charging traffic on
transit links.

In summary, we have shown that a small number of outgoing inter-ISP
connections leads to a major reduction of the overhead and $95$th
percentile up to two order of magnitude. In addition, $4$ outgoing
inter-ISP connections give the minimum attainable overhead of $1$. In
the next section, we explore what is the impact of such a high
reduction on the peers slowdown.

\subsection{Impact on Slowdown}
\label{sec:locality-impact-slowdown}

\begin{figure}[!t]
\centering
\includegraphics[width=1.0\columnwidth]{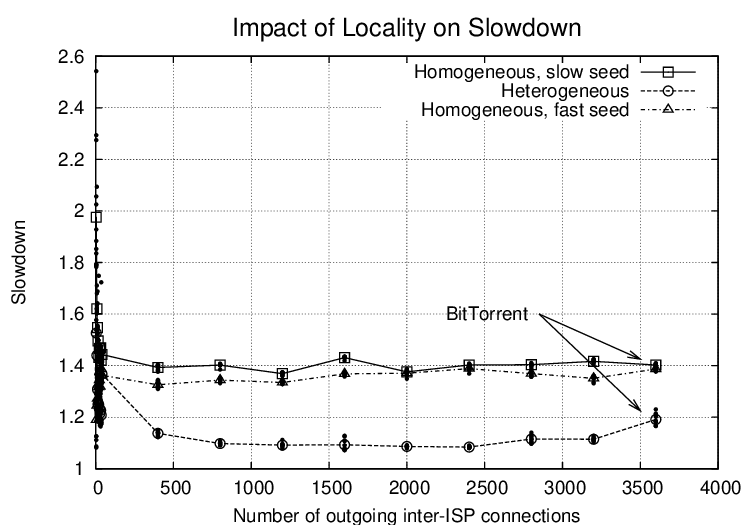}

\caption{\small{Slowdown with $1\,000$ peers and $10$ ISPs. Each
    square, circle and triangle represents the average slowdown on all
    ISPs in a given scenario. Each dot represents this slowdown
    for one ISP.
}}
\label{slowdown-locality}
\end{figure}

The most striking result we observe in Fig.~\ref{slowdown-locality} is
that, whereas for $4$ outgoing inter-ISP connections the overhead is optimal
(only one copy of content uploaded per ISP) and reduced by two orders
of magnitude compared to the \bt{} policy, the slowdown remains
surprisingly low. 

Indeed, Fig.~\ref{slowdown-locality} shows that the number of outgoing
inter-ISP
connections has no significant impact on peers slowdown for the two
scenarios with a fast seed (heterogeneous and homogeneous with a fast
seed) and a negligible impact for more than $16$ outgoing inter-ISP connections
for the homogeneous scenario with a slow seed. This result is
remarkable when one considers the huge saving a small number of
outgoing inter-ISP connections enables for the overhead and $95$th
percentile. 

For the homogeneous scenario with a slow seed, the slowdown increases
by at most $43\%$ for $4$ outgoing inter-ISP connections compared to the case
with the \bt{} policy. This increase is due to a poor piece diversity,
which can be avoided by having a fast initial seed as shown by the two
scenarios with a fast seed in Fig.~\ref{slowdown-locality}. Moreover,
even if a $43\%$ increase is not negligible, it has to be considered
as the worst case. Indeed, as we will show in
section~\ref{bottleneck}, in case of congestion on inter-ISP links,
the slowdown may even improve with a small number of outgoing inter-ISP
connections compared to the \bt{} policy, because that will foster
peers to exchange with peers in the same ISP, thus avoiding congested
paths.

In conclusion, we see that the peer slowdown remains surprisingly low
even for a small number of outgoing inter-ISP connections.

\section{Evaluation of 4 Outgoing Inter-ISP Connections}
\label{small}
We have seen in the previous section that a small number of
outgoing inter-ISP connections dramatically reduces the overhead and $95^{th}$
percentile, and that the slowdown remains low in most cases. 

Whereas this result is encouraging, one may wonder if it is possible
to keep a low overhead and slowdown for a small number of outgoing
inter-ISP connections in more complex scenarios. Therefore, we focus
in the following on $4$ outgoing inter-ISP connections, which leads to
the lowest attainable overhead in our experiments in
section~\ref{sec:locality-impact-overhead}, and we evaluate the
overhead and slowdown when we vary the characteristics of the torrent
(torrent size and number of peers per ISP), or the characteristics of
the network (limitation of the capacity of the inter-ISP links).

Also, as we did not observe a significant impact of the heterogeneous upload
capacity of the peers on our results in section~\ref{locality}, we
consider for the remaining of this paper the homogeneous scenario with
a slow seed. We discuss further the impact of the peers upload
capacity on our results in section~\ref{sec:conclusion}. 

In summary, for this second series of experiments, we consider a
scenario with $4$ outgoing inter-ISP connections, a content of $100$
MB, peers with homogeneous upload capacities, and a slow seed. Then,
we vary the torrent size, the number of peers per ISP, and the
inter-ISP link capacity. We vary only one parameter at a time per
experiment. We consider, in this section, scenarios with the same
number of peers per ISP. Therefore, on average, the number of
incoming inter-ISP connections will be equal to the number of outgoing
inter-ISP connections.

In the following, we do not present results for the $95$th
percentile, as they do not show any significant new insights
compared to the results for the overhead. 

\subsection{Impact of the Torrent Size}
\label{size}
In this section, we make experiments with torrents with $100$, $1\,000$, and
$10\,000$ peers, and $10$ ISPs.

\begin{figure}[!t]
\centering
\includegraphics[width=1.0\columnwidth]{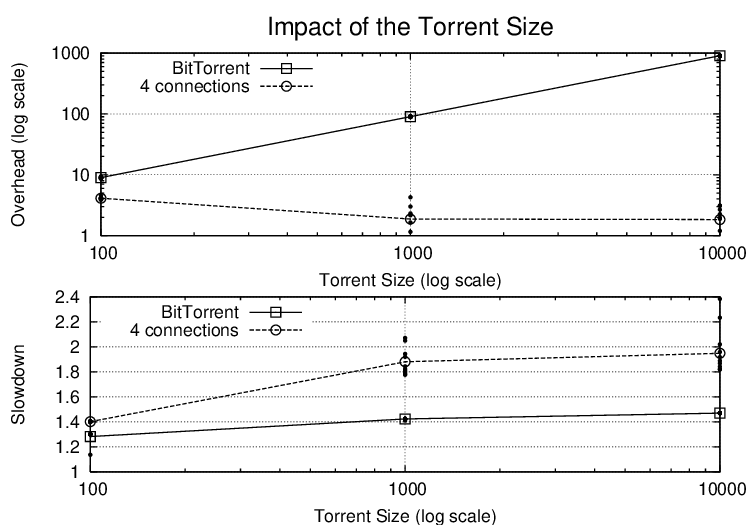}
\caption{\small{Overhead (upper plot) and slowdown (lower plot) for
    torrents with $100$, $1\,000$ and $10\,000$ peers and $10$ ISPs in
    two scenarios: BitTorrent policy, locality policy with $4$
    outgoing inter-ISP connections. Each square and circle
    represents the average overhead (upper plot), or the average
    slowdown (lower plot) for a particular torrent size in a given
    scenario. Each dot represents this overhead (upper plot), or
    slowdown (lower plot) for one ISP. 
}}
\label{overhead-slow-size}
\end{figure}

In Fig.~\ref{overhead-slow-size} upper plot, we see that for a small number of
outgoing inter-ISP connections the overhead is close to one 
independently of the torrent size, whereas for the \bt{} policy it
increases linearly with the torrent size.

For the torrent with $100$ peers, as there are 10 ISPs, there are only
10 peers per ISP. This scenario is interesting because a locality
policy only makes sense when there are enough peers inside each ISP to
be able to keep traffic local. This scenario shows the gain that can
be achieved for a small number of peers per ISP. 
With a torrent of $100$ peers, we save $60\%$ of overhead as compared
to BitTorrent.  With a torrent of $10\,000$ peers, we save
$99.8\%$ of overhead as compared to BitTorrent.

To see the impact of this dramatic overhead reduction on slowdown, we
focus on Fig.~\ref{overhead-slow-size} lower plot. We see that the slowdown
is $8\%$ higher than with the \bt{} policy for a torrent with $100$
peers. For $1\,000$ and $10\,000$ peers, the slowdown is $32\%$ higher
than with the \bt{} policy. 

In summary, we observe that with $4$ outgoing inter-ISP connections, the \bt{}
overhead is optimal and almost independent of the torrent size, which
is at the cost of an increase by around $30\%$ of the slowdown. 

\subsection{Impact of the Number of Peers per ISP}
\label{nr_peers}
In this section, we evaluate $10$, $100$, $1\,000$ and $5\,000$ peers
per ISP. To vary the number of peers per ISP, we vary the number of
ISPs with a constant torrent size of $10\,000$ peers.  Therefore, to
obtain $10$, $100$, $1\,000$ and $5\,000$ peers per ISP, we consider
$1\,000$, $100$, $10$, and $2$ ISPs.

\begin{figure}[!t]
\centering
\includegraphics[width=1.0\columnwidth]{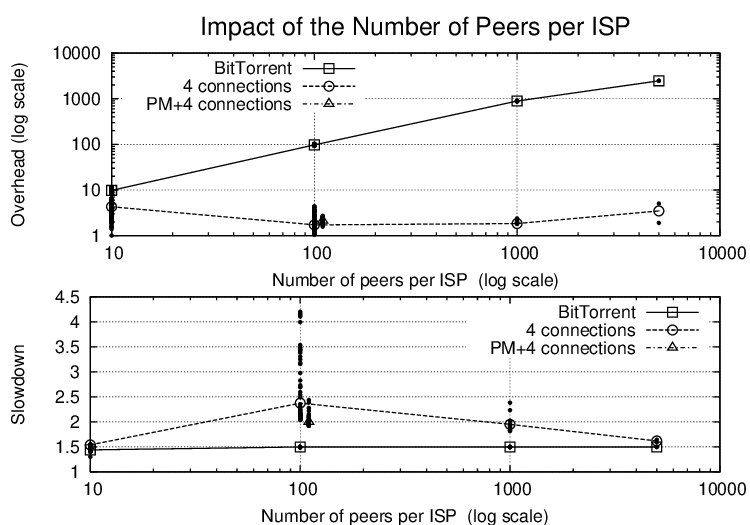}
\caption{\small{Overhead (upper plot) and slowdown (lower plot) with
    $10\,000$ peers and $10$, $100$, $1\,000$, and $5\,000$ peers per
    ISP for two scenarios: BitTorrent policy, locality policy with $4$
    outgoing inter-ISP connections. Each square, circle and triangle
    represents the average overhead (upper plot) or the average
    slowdown (lower plot) for a particular number of peers per ISP in
    a given scenario. Each dot represents this overhead (upper plot)
    or slowdown (lower plot) for one ISP. 
}}
\label{overhead-slow-nr-peers}
\end{figure}

We observe in Fig.~\ref{overhead-slow-nr-peers} lower plot that there
are many outliers points for the scenario with $100$ peers per ISP. In
fact, this scenario is the only one in section~\ref{small} that creates
partitions. Therefore, we also present the result of this experiment
with the Partition Merging (PM) strategy presented in
section~\ref{sec:part-merg-strat}. Indeed, we see that the PM strategy
solves the issue in Fig.~\ref{overhead-slow-nr-peers}. We note that
the results for all the other experiments remain unchanged with the PM
strategy, as they do not create partitions. A detailed evaluation of
the PM strategy is performed in
section~\ref{sec:impact-locality-real}. In the following, we only
consider the results obtained with the PM strategy for the scenario
with $100$ peers per ISP.

Fig.~\ref{overhead-slow-nr-peers} upper plot shows that with $4$
outgoing inter-ISP connections, the overhead remains close to $1$ for any
number of peers per ISP, whereas it increases linearly with the \bt{}
policy. However, this overhead is slightly higher for the scenarios
with $10$ and $5\,000$ peers per ISP.

We also observe on Fig.~\ref{overhead-slow-nr-peers} lower plot that
the slowdown is close to the one of \bt{} for $10$ and $5\,000$ peers
per ISP and around $30\%$ higher than the one of \bt{} for $100$ and
$1\,000$ peers per ISP. This non-monotonic behavior is explained by
the tradeoff that involves two main factors impacting the performance
of \bt{} in this scenario. On the one hand, as the initial seed has a
maximum of $80$ connections to other peers, at most $80$ ISPs can have
a direct connection to the initial seed. All ISPs without direct
connection to the initial seed have to get all the pieces of the
content from other ISPs. Therefore, there is a higher utilization of
the inter-ISP connections and a higher slowdown because the few
inter-ISP connections available to guarantee a high piece diversity
represent a bottleneck. On the other hand, when the number of peers
per ISP decreases, the number of ISPs increases because the torrent
size is constant, thus the global number of inter-ISP connections
increases. Therefore, the overhead increases too, but the slowdown
decreases because there is a sufficient number of inter-ISP connections to
guarantee a high piece diversity.

In summary, we observe that with $4$ outgoing inter-ISP connections, the \bt{}
overhead is optimal and almost independent of the number of peers per
ISP, which is at the cost of an increase by at most $30\%$ of the
slowdown.

\subsection{Impact of the Inter-ISP Link Capacity}
\label{bottleneck}
To explore the impact of inter-ISP link capacity, we consider torrents
with $1\,000$ peers and $10$ ISPs. We vary the inter-ISP link capacity
from $40$kB/s to $100$kB/s by steps of $20$kB/s and from $200$kB/s to
$2\,000$kB/s by steps of $200$kB/s. However, local peers can upload to
their local neighbors (in the same ISP) at $20$kB/s without crossing a
link with limited capacity. For this experiment, all the BitTorrent clients that run
on the same node are located in the same virtual ISP, so limiting the
upload capacity of the node is equivalent to limiting that inter-ISP
link capacity. For an inter-ISP link capacity of $2\,000$kB/s, all the
\bt{} clients that are located on a same node can upload outside this
ISP at their full capacity without any congestion. Therefore, it is
equivalent to the case with no inter-ISP link bottleneck. We use the
tool traffic controller (tc), that is part of the iproute$2$ package,
to limit the upload capacity of each node on which we run
experiments. We deploy our own image of GNU/Linux, on which we have
superuser privileges, on all the nodes we want to limit the upload
capacity. Limiting the upload capacity on each node allows us to
reproduce Internet's bottlenecks in a controlled environment.

\begin{figure}[!t]
\centering
\includegraphics[width=1.0\columnwidth]{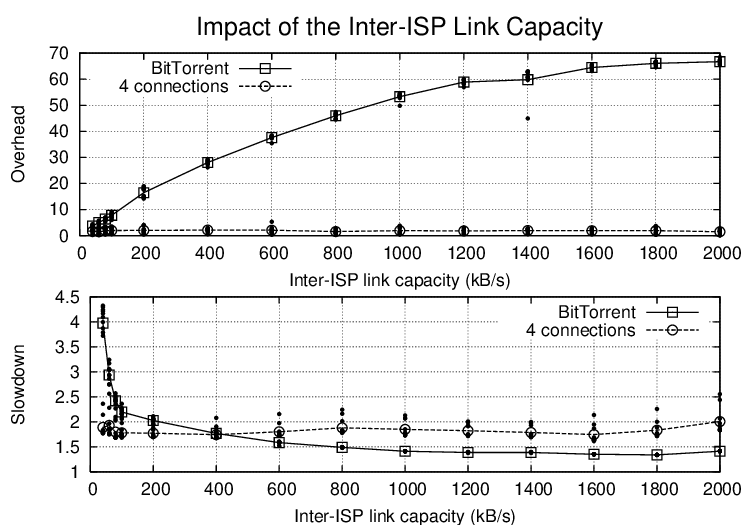}
\caption{\small{Overhead (upper plot) and slowdown (lower plot) with
    $1\,000$ peers and $10$ ISP for various inter-ISP link capacities
    and two scenarios: \bt{} policy, locality policy with $4$ outgoing
    inter-ISP connections. Each square and circle represents
    the average overhead (upper plot) or slowdown (lower plot) for all
    ISPs in a given scenario. Each dot represents this overhead (upper
    plot) or slowdown (lower plot) for one ISP. 
}}
\label{fig:overhead_slow_bottleneck}
\end{figure}

We see in Fig.~\ref{fig:overhead_slow_bottleneck} upper plot that with
$4$ outgoing inter-ISP connections the overhead remains close to $1.5$
for any inter-ISP link capacity. For the \bt{} policy, the overhead
increases with the inter-ISP link capacity.  The reason is that
BitTorrent, due to the choke algorithm, will prefer to exchange data
with local peers when there is congestion on the inter-ISP links,
because those local peers are not on a congested path, thus a larger
\bt{} download speed. For high inter-ISP link capacity, those links
are no more congested, therefore the capacity does not impact anymore
the overhead achieved by the \bt{} policy.

We observe in Fig.~\ref{fig:overhead_slow_bottleneck} lower plot that
with congestion on inter-ISP links, a small number of outgoing
inter-ISP connections improves the peers slowdown. Indeed, for an
inter-ISP link capacity lower than $400$ kB/s, the scenario with the
\bt{} policy becomes slower than the scenario with $4$ outgoing
inter-ISP connections.  The benefit of a small number of outgoing
inter-ISP connections on the slowdown is significant for highly
congested inter-ISP links. For an inter-ISP link capacity of
$40$kB/s, the scenario with with $4$ outgoing inter-ISP connections is
more than $200\%$ faster than with the \bt{} policy.

In summary, the overhead is almost independent of the inter-ISP link
capacity for $4$ outgoing inter-ISP connections, whereas it significantly increases
with the inter-ISP link capacity for the \bt{} policy. 
In addition, when inter-ISP links are congested, we observe a lower
slowdown with the locality policy than with the \bt{} policy. 
We discuss the impact of this result in
the next section. 

\subsection{Discussion}
\label{sec:high-locality-discussion}
We have focused on $4$ outgoing inter-ISP connections and
showed that the overhead is close to $1$ in most scenario and almost
independent of the torrent size, the number of peers per ISP, and the
congestion on inter-ISP links.

But, most surprisingly, the slowdown remains close to the one of the
\bt{} policy in most cases. In some scenarios, the overhead can be
around $30\%$ larger than with the \bt{} policy. Whereas an
increase by $30\%$ cannot be considered negligible, this is a very
positive result for two main reasons. 

First, we remind that our main goal in this section was to minimize
the overhead. We achieved up to three orders of magnitude reduction in
the overhead compared to the \bt{} policy (see
Fig.~\ref{overhead-slow-size} for a torrent with $10\,000$
peers). There is a price to pay for such a huge reduction, which is an
increase by at most $30\%$ in the slowdown. We deem this increase to
be reasonable considering the savings it enables. However, we have
also run experiments with $40$ outgoing inter-ISP connections that are not
shown here due do space limitations, but that are available in a
technical report \cite{LeBlondLegout08_TECH}. We found that with $40$
outgoing inter-ISP connections, the slowdown is always close to the one of
\bt{} at the price of a small increase in the overhead that is close
to $10$ in most of the cases. However, even with this increase in the
overhead, the savings compared to the \bt{} policy are still huge, up
to two orders of magnitude in our experiments. 

Second, the increase we report on the slowdown is the worst one that
can be achieved. Indeed, all our experiments (except the ones
presented in section~\ref{bottleneck}) are performed without
congestion in the network. However, we have shown in
section~\ref{bottleneck} that in case of congestion, our locality
policy can reduce the slowdown compared to the \bt{}
policy. Therefore, on a real network, the slowdown with our locality
policy is likely to be equivalent or even better than the one of the
\bt{} policy.

\section{Real World Scenarios}
\label{sec:real-world-scenarios}
Up to now, we have defined scenarios intended to understand the
evolution of the overhead and slowdown with a small number of
outgoing inter-ISP connections when one varies one parameter at a
time. Those scenarios are not intended to be realistic, but to shed
light on some specific properties achieved with a small number of
outgoing inter-ISP connections. 

In this last series of experiments, we use real world data to build
realistic scenarios. In particular, we will experiment with measured
distribution of the number of peers per AS for real torrents. In the
remaining of this section, we focus on inter-ASes rather than
on inter-ISPs traffic for two reasons. First, the information to perform the
mapping between IP addresses and ASes is publicly available, whereas
there is no standard way to map IP addresses or ASes to ISPs. Second,
ISPs may consist of several ASes. There is no way to find where an ISP
wants to keep traffic local. Indeed, this is most of the time an
administrative decision that depends on peering and transit relations
among its own ASes and the rest of the Internet. However, making the
assumption, as we do, that ISPs want to keep traffic local to ASes is
reasonable, even if there are some cases in which ISPs want to define
locality at a smaller or larger scale than the AS level. Therefore, we
believe that our assumption is enough to give a coarse
approximation of the potential benefits of a small number of outgoing
inter-AS connections at the scale of the Internet.

In the following, we present the crawler we designed to get real
world data. Then we present the results of experiments with real
torrent characteristics. Finally, we give a estimation of the savings
that would have been achieved using our locality policy on all the
torrents we crawled. 

\subsection{Description of the \bt{} Crawler}
\label{sec:descr-bt-crawl}
In order to get real world data, on the $11^{th}$ of
December 2008, we have collected $790\,717$ torrent files on \textit{www.mininova.com} that is
considered one of the largest index of torrent files in the Internet. All those
torrent files were collected during a period of six hours. Out of
these $790\,717$ torrent files, we have removed duplicate ones (around
$1.65\%$ of the files) and all files for torrents that do not have at
least $1$ seed and $1$ leecher. Our final set of torrent files
consists of $214\,443$ files.
 
We have implemented an efficient crawler that takes as input our set
of torrent files and that gives as output the list of the peers in
each of the torrents represented by those files. We identify a peer by
the couple \textit{(IP,port)} where \textit{IP} is the IP address of
the peer and \textit{port} is the port number on which the \bt{}
client of this peer is listening.

Our crawler, which consists of two main tasks, runs on a single server
(Intel Core2 CPU, 4GB of RAM). The first task takes each torrent
file sequentially. It connects first to the tracker
requesting $1\,000$ peers in order to receive the largest number of
peers the tracker can return. Indeed, the tracker returns a number of
peers that is the minimum between the number of peers requested and a
predefined number. The tracker returns a list of $N$ peers, $N$
usually ranging from $50$ to $200$. The tracker also returns the
current number of peers in the torrent. Then, the task computes how
many independent requests $R$ must be performed in order to retrieve
at least $90\%$ of the peers in the torrent when each request results
is $N$ peers retrieved at random from the tracker.

The second task starts a round of $R$ parallel instances of a dummy
\bt{} client, each client started on a different port number, whose
only one goal is to get a list of peers from the tracker. Once a round
is completed, the task removes all duplicates \textit{(IP,port)},
makes sure that indeed $90\%$ of the peers of the torrent were
retrieved, and saves the list of couples \textit{(IP,port)}. In case,
less than $90\%$ of the peers were discovered during the first round,
an additional round is performed. The second task can start many
parallel instances of the dummy \bt{} client for different torrents at
the same time. As the task of the dummy client is simple, we can run
several thousands of those clients at the same time on a single
machine.

At the end of this second task we crawled $214\,443$ torrents
within 12 hours, the largest torrents being crawled in just a few
seconds, and we identified $6\,113\,224$ unique peers.

Finally, we map each of the unique collected peers to the AS it belongs
to using BGP information collected by the RouteViews project
\cite{routeviews}. We found that the unique peers are spread among
$9\,605$ ASes. Even if this way to perform the mapping may suffer
from inaccuracy \cite{MAO04_INFOCOM} \cite{Friedman07}, it is
appropriate for our purpose. Indeed, we do not need to discover AS
relationship or routing information, we just need to find to which AS
each peer belongs to. Even if some mappings are inaccurate, they will
not significantly impact our results, as we consider the global
distribution of peers among all ASes.

This simple but highly efficient crawler enables to capture a
representative snapshot, at the scale of the Internet, of the peers
using \bt{} to share contents the day of our crawl. There are,
however, two limitations to our crawler. First, we only crawled
torrents collected on mininova. Even if mininova is one of the largest
repository for torrent files, it contains few Asian
torrents. Therefore, that means that our results present a lower bound
of the benefit that can be achieved with high locality. Indeed, Asian
torrents are usually large and, due to the geographical locality
inherent to such torrents, spread among fewer ASes than an average
torrent. Therefore, Asian torrents have a larger potential for
locality than other torrents. Second, we are aware that a fraction of
the peers advertised by trackers are fake peers. Indeed, copyright
holders (or representative) join torrents to monitor peers in order to
issue DMCA takedown notices to downloaders \cite{piatekHotSec08}. Also
tracker operators may add fake peers in order pollute the information
gathered by copyright holders. Finally, some peers are identified as
deviant, which means that they do not look like regular peers
\cite{SiganosPam09}.  However, even if the amount of fake peers
accounts for a few percents of the overall peers, considering the large
amount of torrents and peers crawled, we do not believe those fake
peers to significantly bias our results.

\newpage
\subsection{Impact of Locality for a Real Scenario}
\label{sec:impact-locality-real}
In section~\ref{small}, we performed experiments with an homogeneous
number of peers per AS. However, real torrents have an heterogeneous
number of peers per AS, which may adversely impact the overhead
reduction we observed with a small number of outgoing inter-ISP connections.

In order to evaluate the impact of a real distribution of peers per
AS on our experiments, we selected three different torrents from our
crawl with different characteristics. We call those three torrents the
\textit{reference torrents}. The first torrent, that we call
\textit{torrent 1}, is a torrent for a popular movie in English
language. This torrent represents the case of torrents with a worldwide
interest. It has $9\,844$ peers spread among $1\,043$ ASes,
the largest AS consisting of $386$ peers. The second torrent, called
\textit{torrent 2}, is a torrent for a movie in Italian language. This
torrent has $4\,819$ peers spread among $211$ ASes, the largest AS
consisting of $2,415$ peers. This torrent is typical of torrent with
local interest. In particular, this torrent spans less ASes than
\textit{torrent 1}, and the largest AS, belonging to the largest Italian ISP,
represents more than half of the peers of the torrent. The last
torrent, called \textit{torrent 3}, is a torrent for a game. It
has $996$ peers spread among $354$ ASes, the largest AS
consisting of $31$ peers. This torrent is used to evaluate middle
sized torrents with few potential savings with a locality policy, as
there are few peers per AS. 

\subsubsection{Evaluation of  ASes with Heterogeneous Number of Peers}
\label{sec:eval-heter-ases-real}
We have run experiments with the same parameters as the ones of the
homogeneous scenario described in
section~\ref{sec:exper-param-locality}. In particular, we have the
initial seed and all leechers that upload at a maximum rate of 20kB/s,
and a content of $100$ MB. However, we consider scenarios with the
same number of ASes and peers per AS as the three real torrents
considered.  In the following, we focus on experiments performed with
the characteristics of \textit{torrent 1}, as the experiments with the
characteristics of the two other torrents lead to the same
conclusions.

\begin{figure}[!t]
\centering
\includegraphics[width=1.0\columnwidth]{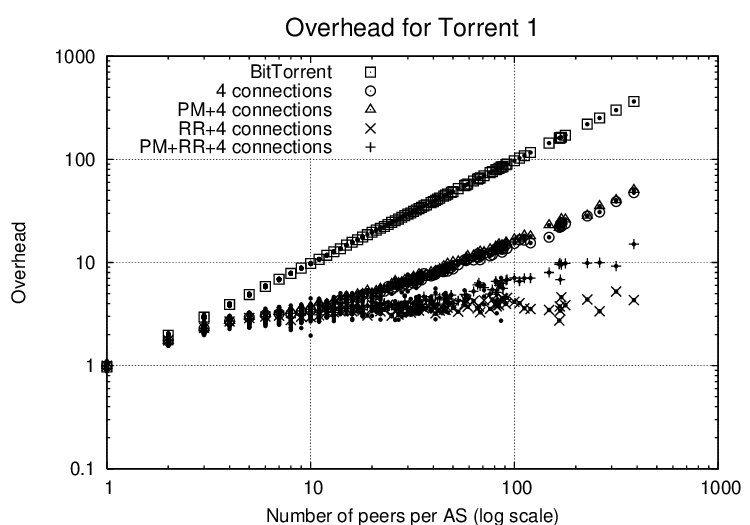}
\caption{\small{Overhead for torrent $1$. Each symbol (rectangle, triangle,
  circle, plus, and cross) represents the average overhead for all ASes
  with the same number of peers for a given scenario. Each dot
  represents the overhead for a single AS. 
}}
\label{sig_Noverhead}
\end{figure}

\begin{figure}[!t]
\centering
\includegraphics[width=1.0\columnwidth]{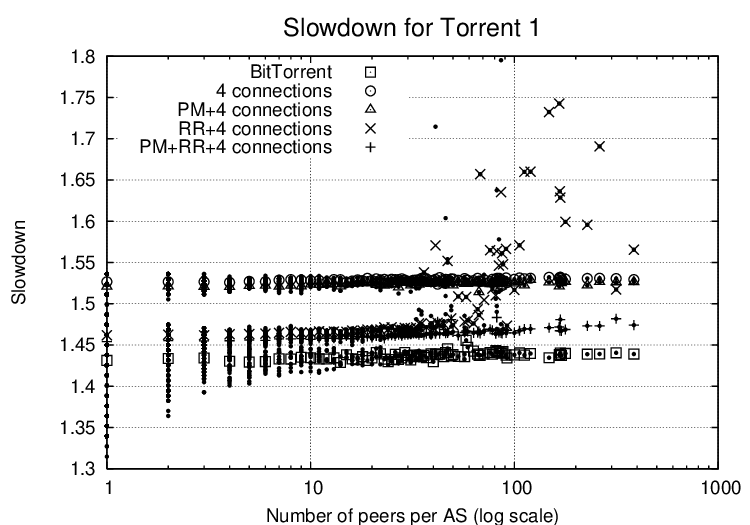}
\caption{\small{Slowdown for torrent $1$. Each symbol (rectangle, triangle,
  circle, plus, and cross) represents the average slowdown
  for all the ASes with the same number of peers and for a given
  scenario. Each dot represents the slowdown for a single
  AS. 
}}
\label{sig_Nslowdown}
\end{figure}

Fig.~\ref{sig_Noverhead} shows the overhead per AS, ordered by number
of peers, for \textit{torrent 1}. As expected, the overhead increases
linearly with the number of peers per AS for the \bt{} policy
(squares).

We observe that the overhead for the scenario with $4$ outgoing
inter-AS connections is one order of magnitude lower than the one of
\bt{} for the largest ASes. However, the overhead is still large for
the largest ASes. In fact, due to the heterogeneity in the number of
peers per AS, as explained in section~\ref{sec:round-robin-strategy},
the largest AS will have more incoming inter-AS connections than small
ones. Therefore, large ASes will have a larger number of
inter-AS connections, thus a larger overhead than small ASes.

The solution to this problem is to use the Round Robin (RR) strategy
introduced in section~\ref{sec:round-robin-strategy}. Indeed,
Fig.~\ref{sig_Noverhead} shows that the overhead is significantly
reduced with the RR strategy (cross). However, we see in
Fig.~\ref{sig_Nslowdown} that the slowdown for the largest ASes
increases significantly compared to the other scenarios. Indeed, as the
RR strategy spreads uniformly the incoming inter-AS connections on all
ASes, each AS will have on average $8$ inter-AS connections in total
($4$ outgoing and $4$ incoming). Therefore, for the largest ASes, only
few peers will have an inter-AS connection. Once those peers leave
the torrent after their completion, the largest AS will become
partitioned with a large number of peers waiting for new pieces from
the initial seed. Thus, a larger slowdown.

To solve this issue, we made experiments with the Partition Merging
(PM) strategy that is supposed to repair partitions quickly (see
section~\ref{sec:part-merg-strat}). Indeed, we see in
Fig.~\ref{sig_Nslowdown} that the scenario with $4$ inter-AS outgoing
connections and the PM+RR strategies (plus) gives the best slowdown
over all the scenario using a locality policy, close to the one of the
\bt{} policy. This significant improvement is at the cost of a small
increase in the overhead, see Fig.~\ref{sig_Noverhead} (plus), but the
overhead remains up to two orders of magnitude lower than with the
\bt{} policy.

To show that the PM strategy does not impact our results when there is
no partition, we consider a scenario with $4$ outgoing inter-AS
connections and the PM strategy only. We see in
Fig.~\ref{sig_Noverhead} that the overhead of this scenario (triangle) is almost
indistinguishable from the scenario without the PM strategy
(circle). We observe in Fig.~\ref{sig_Nslowdown} that the slowdown for
both scenarios is also indistinguishable. Therefore, the PM strategy
does not bias our results by artificially increasing the number of
inter-AS connections. 

In summary, the PM+RR strategies solve issues with real torrents and
enable huge overhead reduction and a low slowdown.

\subsubsection{Evaluation of Churn}
\label{sec:evaluation-churn-real}
In this section, we run all our experiments with the characteristics
of \textit{torrent 1}. In particular, we consider scenarios with the same
number of ASes and peers per AS as \textit{torrent 1}.  To evaluate the
impact of churn, we start a first set of $9\,844$ peers uniformly
within the first $60$ seconds in a first experiment, and within the
first $6\,000$ seconds in a second experiment.  Then, when each of
those peers completes its download, we start a new peer from a second
set of $9\,844$ peers. Hence, we model the three phases of a real
torrent's life: flashcrowd, steady phase, and end phase
\cite{PAM04_BT}. The first phase, the flashcrowd, occurs while all
peers of the first set join the torrent. The second phase, the steady
phase, occurs when the number of peers in the torrent remains
constant. This is when peers in the first set start to complete and
that peers in the second set are started to replace those peers in
order to keep the torrent size constant to $9\,844$ peers. The last
phase, the end phase, occurs at the end of the torrent's life, when
the last peers complete their download and no new peer joins the
torrent. This is when there is no more peers in the second set to
compensate departure of peers.

\begin{figure}[!t]
\centering
\includegraphics[width=1.0\columnwidth]{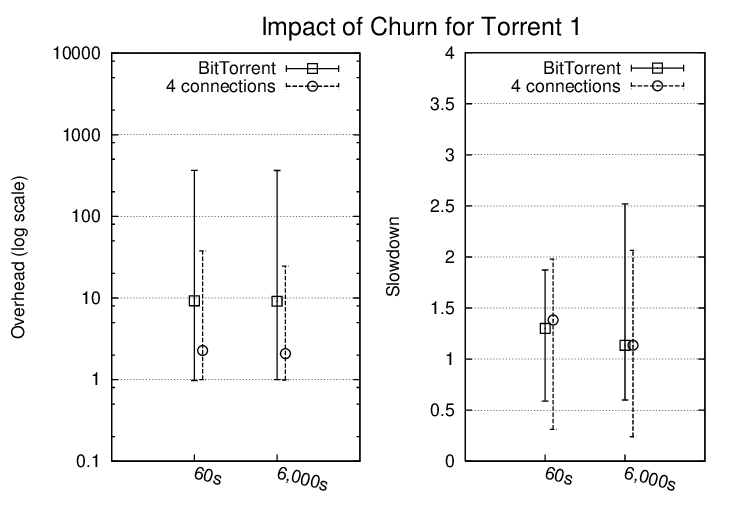}
\caption{\small{Overhead (left plot) and slowdown (right plot) with churn of
  $60$s or $6000$s for torrent $1$ in two
  scenarios: BitTorrent policy, locality policy with $4$ outgoing
  inter-AS connections. Each square and circle represents the average
  overhead (left plot) or average slowdown (right plot) on all ASes
  for a specific scenario. The error bars represent the minimum and
  maximum overhead on all ASes (left plot), and the minimum and
  maximum slowdown on all peers (right plot). 
}}
\label{fig:sig_churnReference}
\end{figure}

Large torrents, like \textit{torrent 1}, represent the most
challenging scenario in case of churn. Indeed, small torrents will
have just one to a few peers per AS. Therefore, as most connections
among peers will be inter-AS, the locality policy will not
significantly constrain the peers connectivity graph. Consequently, this
graph will be random, unlike with a large torrent whose graph
is clustered per AS, thus a better robustness to AS isolation in case
of churn.

We see in Fig.~\ref{fig:sig_churnReference} left plot that the maximum overhead
is reduced by one order of magnitude with $4$ outgoing inter-AS
connections compared to the \bt{} policy. Moreover, this reduction has
no negative impact on the slowdown as shown by
Fig.~\ref{fig:sig_churnReference} right plot. 

In summary, even with churn the overhead is reduced and the slowdown
remains low independently of the churn period with $4$ outgoing
inter-AS connections.  We also run experiments with a churn of $600$
seconds, with $10$ to $1\,000$ homogeneous ASes, and other arrival
patterns \cite{LeBlondLegout08_TECH} without any significant impact on
our conclusions.

\subsection{Estimation of Locality Benefits at the Scale of the
  Internet}
\label{sec:estim-local-benef-real}
In this section, we want to estimate the benefits our locality policy
would have had on the torrents we crawled. In our crawl, $117\,677$
torrents and $6\,643$ ASes cannot benefit from a locality policy,
because there is at most one peer per AS per torrent. However, we want
to show that despite most of the torrents and ASes cannot benefit from
a locality policy, the implementation of a locality policy at the
scale of the Internet would be highly beneficial.

In order to make the estimation of the benefits of our locality policy,
we make several assumptions. First, we estimate the inter-AS traffic
in all the torrents we crawled by assuming that all the peers we found
start downloading the content at the same time and stay connected to
the torrent for the entire duration of their download. Indeed, we have
not captured temporal information, which means that we do not know how
long each peer stayed in each torrent. However, it is hard to know if
we underestimate or overestimate the potential for locality of those
peers. Indeed, for torrents in a flash crowd phase, most peers are
leechers and the population increases with time. For those torrents,
we are likely to underestimate the benefits of our locality
policy. For torrents in an end phase, most peers are seeds and the
population is decreasing, therefore, it is likely that we overestimate
the benefits of our locality policy. We believe our assumption
to be reasonable and to provide, on average, at least a coarse
estimation of the inter-AS traffic generated by all the peers we
crawled.

Second, we assume that peers have the same probability to exchange
data with any peer in its peer set. Therefore, we assume that peers
have the same upload capacity and that there is no network bottleneck
that bias the peer selection with the choke algorithm. Here again, it
is hard to assess the exact impact of this assumption on the accuracy
of our results, but we believe that, considering the large number of
torrents we crawled, our estimation of the inter-AS traffic is
reasonable.

\begin{figure}[!t]
\centering
\includegraphics[width=1.0\columnwidth]{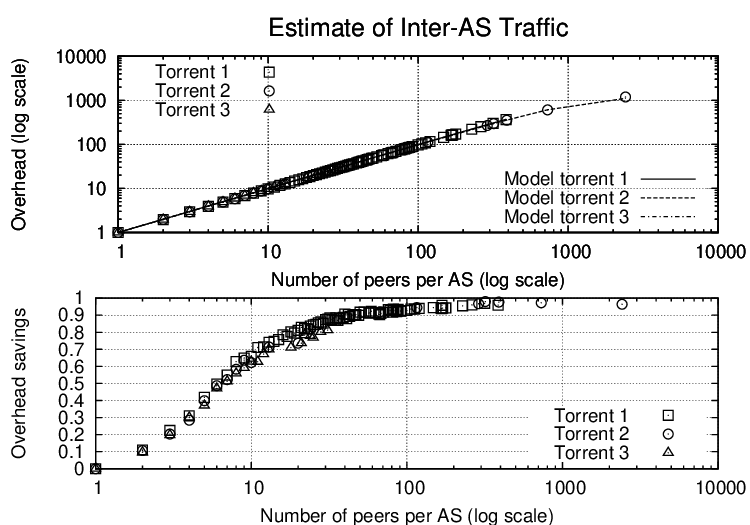}
\caption{\small{Overhead for the three reference torrents with the
    \bt{} policy fitted with the estimation of this overhead using a
    simple model (upper plot), and overhead savings with $4$ outgoing
    inter-AS connections with PM+RR compared to \bt{} policy for the
    three reference torrents (lower plot). 
}}
\label{fig:sig_savings}
\end{figure}

In order to estimate the benefits of our locality policy, we first
estimate the inter-AS traffic generated with the \bt{} policy, then we
estimate the overhead savings enabled by our locality policy. 

To estimate the inter-AS traffic generated by the torrents we assume
that the probability that a peer in a given AS will upload data to a
peer in another AS is only a function of the number of inter-AS
connections of the ASes. In particular, for a torrent of size $S_T$, an
AS $A$ of size $S_{A}$, and a content of size $C$, the inter-AS
traffic uploaded from $A$ is $(1-\frac{S_{A}}{S_T}) \cdot S_{A} \cdot
C$. While this model is simple, we see in Fig.~\ref{fig:sig_savings}
upper plot that it matches well the inter-AS traffic uploaded from
each AS that we measured for the three reference torrents.
Then, for each AS and each torrent, we compute using the simple model
the inter-AS traffic.

To estimate the inter-AS traffic generated by the torrents we crawled
with the locality policy with PM+RR, we use the overhead savings we
obtained with experiments with the three reference torrents. Indeed,
we see in Fig.~\ref{fig:sig_savings} lower plot, that the overhead
savings of our locality policy with PM+RR compared to the \bt{} policy
depends on the number of peers per AS, but not on the torrent
size. Therefore, we use the average overhead savings computed on the
three reference torrents for each AS size to compute the reduction of
inter-AS traffic. We also made the same experiments without the PM+RR
strategy to estimate the inter-AS traffic with our locality policy
without those strategies, and we observed that the savings depend on
the number of peers per AS, as well. 

Fig.~\ref{fig:sig_savings} lower plot shows that even with a small number of
peers per AS, the overhead savings are already high. For instance, with
$5$ peers per AS, the overhead with our locality policy is $40\%$
lower than the one with the \bt{} policy. 

\label{sec:proj-local-benef}
\begin{figure}[!t] 
\centering
\includegraphics[width=1.0\columnwidth]{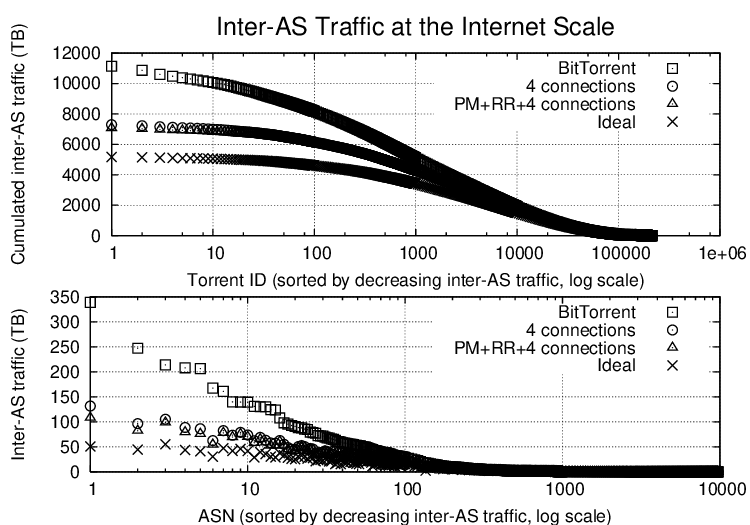}
\caption{\small{Estimation of the cumulated inter-AS traffic for all torrents in terabytes
  (upper plot) and inter-AS traffic per AS in terabytes (lower
  plot). 
}}
\label{fig:sig_traffic}
\end{figure}

Now, we focus on the impact of those savings at the scale of all the
torrents we crawled. We see in Fig.~\ref{fig:sig_traffic} upper plot
the cumulative inter-AS traffic for each torrent we crawled. The $100$
(resp. $10\,000$) largest torrents generate $26\%$ (resp. $82\%$) of
the inter-AS traffic. The ideal policy corresponds to the inter-AS
traffic generated when only one copy of the content is uploaded per AS
and per torrent. We see that the cumulative inter-AS traffic with the
\bt{} policy is $11.6$ petabytes, and that with $4$ outgoing inter-AS
connections it is $7.3$ petabytes (and $7$ petabytes with the PM+RR
strategies), which is only $41\%$ larger ($35\%$ with PM+RR) than what
the ideal policy achieves. Therefore, our locality policy enables a
significant reduction of the inter-AS traffic at the scale of the
Internet.

The $50$ (resp. $300$) largest ASes represent $45\%$ (resp. $84\%$) of
the total inter-AS traffic.  Interestingly, we see in
Fig.~\ref{fig:sig_traffic} lower plot, that the ASes with the largest
inter-AS traffic are also the ones that benefit from the most
significant inter-AS traffic reduction with our locality policy. We
checked manually the $50$ largest ASes to make sure that they do not
belong to copyright holders (or piracy tracking companies) to be sure
that most of the peers in those ASes are real peers
\cite{SiganosPam09}.

In summary, a high locality policy can reduce by up to $40\%$ the
inter-AS traffic for the $214\,443$ real torrents we crawled spread
across $9\,605$ ASes.

\section{Related Work}
\label{work}

Karagiannis et al. \cite{kara05_IMC} first introduced the notion of
locality in the context of \p2p{} content replication. They show
monitoring the access link of an edge network and running simulations
using a log collected from a \bt{} tracker for a single torrent
\cite{PAM04_BT} that peer-assisted locality distribution is an
efficient solution for both the ISPs and the end-users. 

P4P \cite{p4p} is a project whose aim is to provide a light-weight
infrastructure to allow cooperation between \p2p{} applications and
ISPs.  Xie et al. presented small scale experiments (with
between $53$ and $160$ PlanetLab nodes) on two specific scenarios. They
also reported on a field test experiment around $60\%$ of inter-ISP
traffic savings with P4P for a single ISP and a single large torrent.

Aggarwal et al. \cite{agga07_CCR} present an architecture that is
similar by some aspects to P4P. The authors define the notion of
\textit{oracle} that are supplied by ISPs in order to propose a list
of neighbors to peers. They perform their evaluation on Gnutella using
simulations and small scale experiments with 45 Gnutella nodes.

Another approach that requires no dedicated infrastructure is Ono
\cite{ono}. Ono clusters users based on the assumption that clients
redirected to a same CDN server are close. The authors have developed
an Ono plugin for the Vuze client. The authors reported measurement
results collected from $120\,000$ users of the Ono plugin over a $10$
month period. They reported up to $207\%$ performance increase in
average peer download completion time. However, the authors did
not give an explicit inter-ISP traffic reduction, but showed a
reduction of the path length between peers in terms of IP and AS hops.

Bindal et al.~\cite{bind06_ICDCS} present the impact of a
deterministic locality policy on ISPs' peering links load and on
end-users experience. The authors considered simulations on a scenario
with $14$ ISPs with $50$ peers each, thus a torrent of $700$ peers.

Our work significantly differs from those previous ones, by being the
first one to extensively evaluate the impact of key parameters like
the number of inter-ISP connections, the torrent size, the
distribution of peers per ISP, the inter-ISP bottlenecks, the churn
rate, and the peers upload capacity using large scale experiments and
real world data. In particular, we considered $214\,443$ real torrents
spread across $9\,605$ ASes (it was a single large torrent and a
single AS for the P4P field test \cite{p4p}) and showed that using
only four inter-ISP connections (it was $20\%$ of inter-ISP
connections for the P4P field tests) we can reduce the inter-ISP
traffic at the scale of the Internet by $40\%$.

\section{Discussion}
\label{sec:conclusion}
Our work is intended to be complimentary to previous works
\cite{kara05_IMC,p4p,ono} by answering the two fundamental questions:
How far can we push \bt{} locality? What is at the scale of the
Internet the reduction of inter-ISP traffic that can be achieve with
locality?

In this paper, we have performed an extensive evaluation of the impact
of a small number of inter-ISP connections on overhead and
slowdown. We have run experiments with up to $10\,000$ real \bt{}
clients in a variety of scenarios, including scenarios based on real
data crawled from $214\,443$ torrents representing $6\,113\,224$
unique peers spread among $9\,605$ ASes.

Our main findings are that a small number of inter-ISP connections
will dramatically reduce the overhead and keep the slowdown low
independently of the torrent size, the number of peers per ISP, the
upload capacity of peers, or the churn. We have introduced two new
strategies called Round Robin and Partition Merging that make the use
of a small number of inter-ISP connections feasible for real torrents
of the Internet. 

However, we do not advocate for such small number of inter-ISP
connections in real deployments. Instead, we intend to increase
confidence in \bt{} locality by showing that even in case of high
locality \bt{} still performs extremely well, and that with high
locality the inter-ISP traffic reduction can be up to $40\%$ on the
torrents we crawled, which is $4.6$ petabytes of data.

Finally, we have explored, in section~\ref{locality}, a scenario with
three classes of upload capacity spread uniformly over all peers. We
have shown that the results obtained for this scenario do not
significantly differ from an homogeneous scenario. However, we did not
explored scenarios with realistic peers upload capacity
distribution. In fact, it is hard, if not impossible, to obtain this
information at the scale of the Internet. Moreover, we believe that
the impact of the real heterogeneity of peers will be better explored
with a real deployment. Our work shows that a real deployment makes
sense, and we are currently working with BitTorrent inc. to implement,
evaluate, and possibly deploy a locality policy in the uTorrent
client, the most popular \bt{} client with more than $40$ millions
users.

\end{document}